\documentclass[%
 reprint,
showpacs,preprintnumbers,
 aps,
]{revtex4-1}

\usepackage{lineno,hyperref}

\usepackage{dcolumn}

\usepackage{float}
\usepackage{bm}
\usepackage{alltt}
\usepackage{listings}
\usepackage[all]{xy}
\usepackage{tikz}
\usetikzlibrary{arrows}

\usepackage{geometry}

\usepackage{latexsym,enumerate,amsmath,amssymb}
\usepackage{amsthm}
\usepackage{verbatim}
\usepackage{eufrak}

\def\eg{{\it e.g.}}

\usepackage{graphicx}
\usepackage{hyperref}
\usepackage{xcolor}
\usepackage{multirow}

\newcommand{\wt}[1]{\widetilde{#1}}

\begin{document}
\preprint{APS/123-QED}

\title{Nonlinear Constitutive Models for Nano-scale Heat Conduction}

\author{Weiqi Chu}
\email{wzc122@psu.edu}
\affiliation{Department of Mathematics, the Pennsylvania State University, University Park, PA 16802, USA.}%

\author{ Xiantao Li}
\email{xli@math.psu.edu}
\affiliation{Department of Mathematics, the Pennsylvania State University, University Park, PA 16802, USA.}%


\begin{abstract}
{ 
We present a rigorous approach that leads, from a many-particle description, to a nonlinear,  stochastic constitutive relation for the modeling of  transient heat conduction processes at nanoscale.  By enforcing statistical consistency, in that the statistics of the local energy is consistent
with that from an all-atom description, we identify the driving force as well as the model parameters in these generalized constitutive models. 
} 
\end{abstract}

\maketitle

Heat conduction properties of nano-mechanical systems have  significant impacts on the performances of nano devices. 
There have been enormous experimental and numerical observations that indicate the breakdown of the conventional model of heat conduction, the Fourier's Law. The indications include, but not limited to, the size dependence of the heat conductivity, propagation heat pulses behavior, and delay phenomena \cite{alvarez2007memory,ChOk-2008}.  

A remarkable approach to model non-Fourier behavior is the Cattaneo and Vernotte (CV) model \cite{cattaneo1958form,vernotte1958paradoxes}, which  eliminates the paradox of infinite speed of the temperature propagation. 
Further generalizations, e.g., the Guyer-Krumhansl (GK) model \cite{guyer1966solution},  Tzou model  \cite{Tzou-2001},
the extended thermodynamics models \cite{alvarez2007memory}, involve auxiliary equations for the heat flux. Another extension is nonlinear heat conduction model, where the heat conductivity is temperature-dependent. The nonlinearity gives rise to  traveling waves, which also implies that temperature can propagate with finite speed \cite{pascal1992nonlinear}. 

On the other hand, the availability of molecular dynamics (MD) models has encouraged a great deal of effort to simulate heat conduction
 directly at the nano-mechanical scale. Despite the great amount of computational results, 
 the atomic-scale origin of generalized heat conduction models is still not fully understood. For example, it is well known that 
 the heat conductivity in the conventional Fourier's law can be expressed using the Green-Kubo formula \cite{Toda-Kubo-2}. But how the parameters in the CV and GK models can be related to the statistics of underlying atomic trajectories have not been established. Several important questions remain, including: In general, what is the driving force  for  energy transport? How does one describe  processes occurring at the transient stage? Is there a systematic procedure to identify model parameters?  Is it possible to quantify the uncertainty associated with generalized heat conduction models? 

This paper presents a first-principle based derivation of generalized, nonlinear, stochastic  heat conduction models, obtained directly from 
a full MD model, aiming to address aforementioned issues.  The stochastic nature plays three roles: (1) It identifies the driving force for heat conduction, which can be nonlinear; (2) It enables a parameter identification through the underlying statistics; (3) It leads to stochastic solutions for which one can quantify the uncertainty. In addition, it will be shown that the CV model, along with its extensions, can be derived. The model parameters can all be linked directly to the statistics of the local energy.

\begin{figure}[H]
\begin{center}
\includegraphics[width=0.4\textwidth]{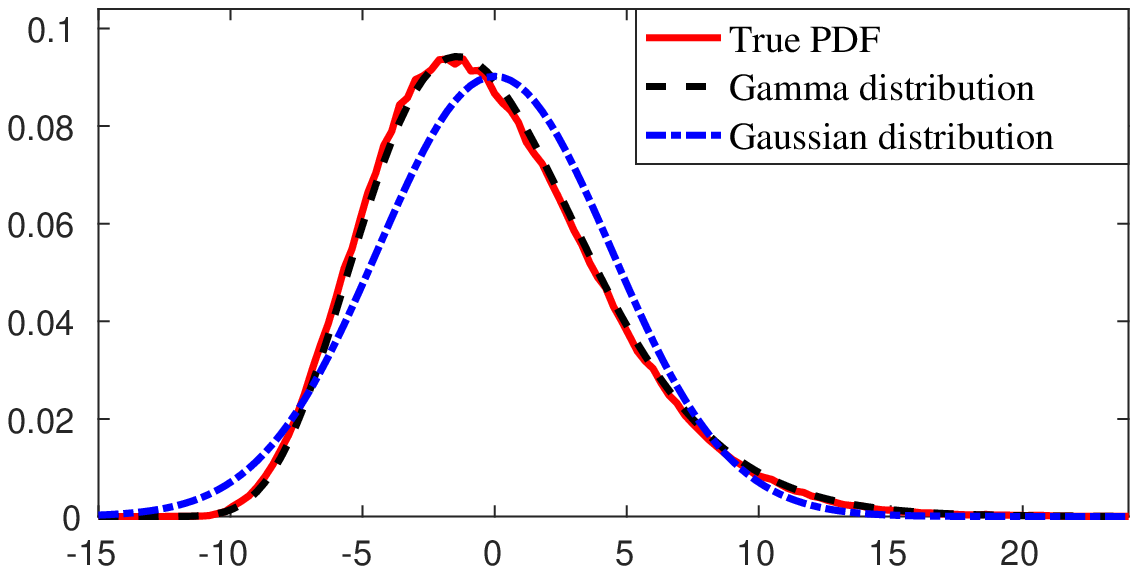}
\includegraphics[width=0.4\textwidth]{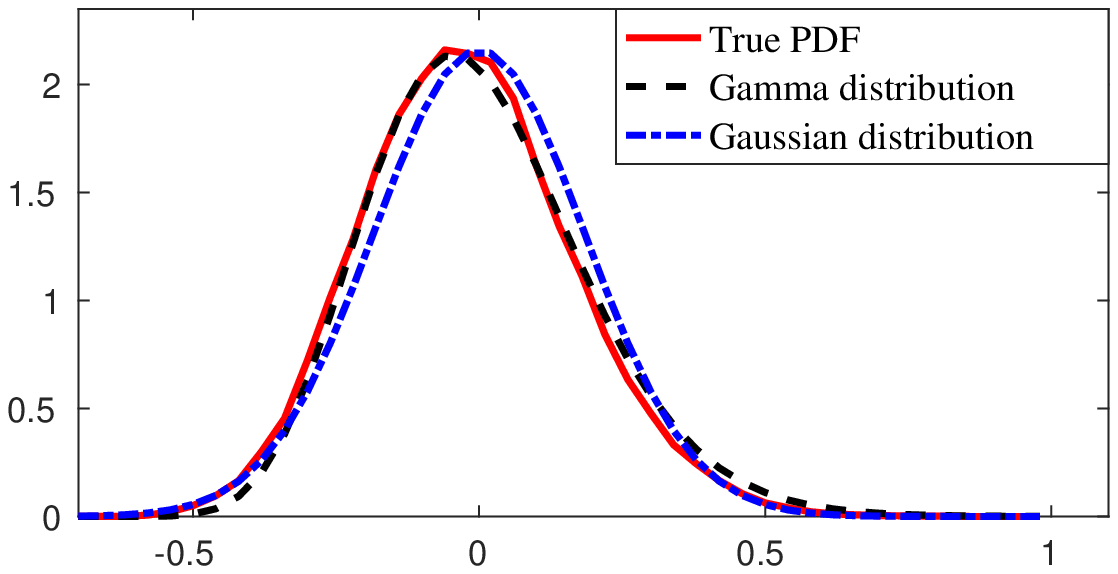}
\caption{The PDF of the local energy at equilibrium, compared with Gaussian and Gamma distributions.
Left: a 1D chain; Right: a nanotube.}
\label{fig: 1d-pdf}
\end{center}
\end{figure}

 We start with observations from some numerical experiments. The first example is a 1D Fermi-Pasta-Ulam (FPU) model, studied extensively in  statistical physics  \cite{lepri2003thermal}. The 1D chain is divided into blocks, each containing 10 atoms. The second example is a carbon nano-tube, with interactions modeled by the Tersoff potential \cite{Te86}. The system  is again  divided into blocks in the longitudinal direction, each containing 16 atoms, with $length$ denoted by $h$ . In both examples, the time series of the local energy in each block is generated from a direct MD simulation under the canonical ensemble. The probability density function (PDF) of the local energy (denoted by $a(t)$), is shown in Fig. \ref{fig: 1d-pdf}.

An interesting finding is that the statistics of the energy is non-Gaussian. The PDF actually fits better to a shifted Gamma distribution,
\begin{equation}
\rho(\xi) \propto (\xi+\mu)^{\alpha-1}\exp\left[-\beta(\xi+\mu)\right].
\end{equation}

We also observed that the local energy among the blocks are independent. So a reasonable ansatz for the joint PDF $\rho(a)$ is
\begin{equation}
\rho \propto \prod_{i=1}^{n_{block}} (a_{i}+\mu)^{\alpha-1}e^{-\beta(a_{i}+\mu)},
\label{eq: rhoa}
\end{equation}


To ensure the consistency with the true statistics, we first write this density as, 
\(
\rho = \exp[-W(a)]/\int \exp[-W(a)] da.
\)
We then define $b(t)$ as the driving force of conduction,
\begin{equation}
b(t) \overset{def}{=} -\frac{\delta W(a)}{\delta a}.
\end{equation}

We extend the Mori-Zwanzig (MZ) procedure \cite{Mori65,Zwanzig73} by defining the projection,
\(\mathcal{P}{f} \overset{def}{=} \langle {f},{b}\rangle \langle {b},{b}\rangle ^{-1}{b}.\)
The formalism yields a generalized Langevin equation (GLE), 
\begin{equation}
\dot{a}(t) = \int_0^t \theta(t-s) {b}(s) ds + F(t),
\label{eq: oblq}
\end{equation}
where  $\theta(s)=\langle \mathcal{L}F(s),{b} \rangle  \langle {b},{b}\rangle ^{-1}$ with $F(t)=e^{t\mathcal{QL}}\mathcal{QL}{a}$, $\mathcal{Q}=I-\mathcal{P},$ and $\mathcal{L}$
is the Liouvilian. 

%
%
%
%
%
%
%
\medskip
The GLE is  nonlocal, which looks quite different from the conventional and extended models. To draw connections, we approximate the kernel function via its Laplace transform:
\begin{equation}
{\Theta}(\lambda) ~ \overset{def}{=} ~ \int_0^{+\infty} \theta(t) e^{-t/\lambda} dt.
\end{equation}

We denote  the statistics of the energy by,
\begin{equation}
 M(t)=\langle {a}(t),{a}^\intercal\rangle, \text{and}\;\; N(t) =\langle {b}(t),{a}^\intercal\rangle. 
\end{equation}

We assume that the statistics can be extracted from, \eg, equilibrium MD simulations. We define short-time statistics, 
\begin{equation}
M_\ell=\langle  {a}^{(\ell)} (0), {a}^\intercal \rangle,\quad
N_\ell=\langle {b}^{(\ell)} (0), {a}^\intercal \rangle.
\end{equation}
$\wt{M}(\lambda)$ and $\wt{N}(\lambda)$ can be expanded around $0_+$,
\begin{equation}\label{eq: MN}
\begin{aligned}
&\wt{M}(\lambda) = \lambda M_{0} + \lambda^{2} M_{1} + \lambda^{3} M_{2} + \cdots,\\
&\wt{N}(\lambda) = \lambda N_{0} + \lambda^{2} N_{1} + \lambda^{3} N_{2} + \cdots.
\end{aligned}\end{equation}

For long time statistics, we define,
\begin{equation}
N_{\infty}=\lim_{\varepsilon\to 0+}\int_0^{+\infty} e^{-\varepsilon t} N(t) dt.
\end{equation}

One systematic procedure for reducing the GLE  is  the embedding technique \cite{lei2016data,ma2016derivation}, which approximates $\Theta(\lambda)$  by rational functions,
\[ R_{k,k}= \big[I-\lambda B_1 - \cdots - \lambda^k B_{k}\big]^{-1}
   \big[\lambda A_1 +  \cdots + \lambda^k A_{k}\big].\]

Assuming that ${a}$ is uncorrelated with the noise term, we multiply the equation \eqref{eq: oblq} by the transpose of ${a}$ and take Laplace transform. 
With $\Theta$ approximated by $R_{k,k}$, we obtain, 
\begin{equation}\label{eq: interp}
\frac{1}{\lambda}\wt{M}(\lambda) - M_{0}= R_{k,k}(\lambda) \wt{N}(\lambda).
\end{equation}
Now the coefficients in the rational function can be determined by combining Eq. \eqref{eq: MN} and Eq. \eqref{eq: interp}, by matching the coefficients in the expansion. For example, when $k=1, $ we find that $A_1=M_2N_0^{-1}.$ In general, one can match the first $2k-1$ coefficients, yielding $2k-1$ linear equations for the coefficients $A_i$'s and $B_i$'s. The remaining condition is imposed at $\lambda \to +\infty,$ which incorporates long-time statistics, yielding  $ M_{0} = B_{k}^{-1}A_{k}N_{\infty}.$ As it turns out, without the long-time statistics, the  resulting model will be wave-type of equations, with no dissipation. 


Thanks to the rational approximation, the resulting reduced model can be each converted back to the time domain, and expressed as a set of differential equations. The memory term is embedded in an extended system {\it without} memory. 
Here we present the models for $k=0, 1,$ and $2$, which will be referred to as zeroth, first and second order models, respectively.
 The random noise term in the GLE \eqref{eq: oblq} will be approximated indirectly by introducing Gaussian noises in the extended system , in such a way that the statistics of the local energy is consistent. 


%
%




\noindent{\bf Zeroth order model.}  For $k=0,$ $\Theta$ is approximated by a constant matrix, 
\begin{equation}
 \Theta \approx -M_0 N_\infty^{-1}
\end{equation}
which corresponds to a delta function in the time domain.  Numerical results suggest that this matrix is tri-diagonal, corresponding to a 
discrete Laplacian operator, $ \Theta \approx -  \kappa\nabla_h^2. $ Here $ \nabla_h a_j \overset{def}{=} (a_{j+1}-a_j)/h$ and  $ \nabla_h^2 a_j \overset{def}{=} (a_{j+1}-2a_j+a_{j-1})/h^2$. Therefore, the GLE is reduced to, 
\begin{equation}
  \dot{{a}} = \nabla_h^2   \kappa \frac{\delta W({a})}{\delta a}  + \sigma \xi(t).
\label{eq: zero_GLE}  
\end{equation}
The selection of $\sigma$ is standard: if  $\sigma \sigma^\intercal  = -2\kappa   \nabla_h^2 $,
then  \eqref{eq: zero_GLE}   has the correct  PDF \eqref{eq: rhoa}. Eq. \eqref{eq: zero_GLE} can be viewed as a discretized nonlinear heat equation with space-time white noise. Very interestingly, this coincides with the stochastic phase-field crystal model \cite{berry2008melting}.

The corresponding stochastic constitutive relation for the heat current is thus given by,
\begin{equation}
  q = - \kappa \nabla_h \frac{\delta W({a})}{\delta a} + \sqrt{2 \kappa}  \xi(t),
\end{equation}
which is a nonlinear, stochastic generalization of the Fourier's Law.

%

\noindent{\bf First order model.}  
When $k=1,$
\( R_{1,1}(\lambda) = \big[I - \lambda B_1]^{-1} \lambda A_1.\)
In the time-domain, if we define 
${z}= \int_{0}^{t} \theta(t-s){b}(s) ds,$ 
then we can write down an auxiliary equation,
 $\dot{{z}}= A_1 {b} + B_1 {z}.$
It can be shown that the interpolation conditions lead to  $A_1=-M_2=- \gamma_1 \nabla_h^2$, and $B_1= -\gamma_1/\kappa I.$ The GLE \eqref{eq: oblq} is simplified to,
\begin{equation}
\left\{\begin{aligned}
   \dot{{a}} =&  {z}, \\ 
   \dot{{z}} = & \nabla_h^2   \gamma_1 \frac{\delta W({a})}{\delta a}  - \frac{\gamma_1}\kappa {z} + \sigma \xi,
\end{aligned}\right.
\label{eq: firstmodel}
\end{equation}

By choosing  $\sigma\sigma^{\intercal}=-2\gamma_1 \nabla_h^2$, the first order model has the correct equilibrium PDF, 
\begin{equation}
{ \rho(a,z) = \exp -\left(W(a) +\frac{\gamma_{1}}{2\kappa^{2}}z^\intercal \nabla_h^{-2} z\right) .}
\end{equation}

The corresponding constitutive relation for the heat flux can be written as,
\begin{equation}
 \tau_{1} \dot{q} + q  = -\nabla_h  \kappa \frac{\delta W({a})}{\delta a} + \sqrt{2\kappa} \xi(t).
\end{equation}
This is an interesting generalization of the CV model \cite{cattaneo1958form}. Not only have we identified the origin of the relaxation parameter,
$\tau_{1}= \kappa/\gamma_1$, we also incorporated a nonlinear driving force and a stochastic noise. 

The model can also be written as 
\begin{equation}
\tau_{1} a_{tt}  +  a_{t}  =  \nabla_h^2 \kappa  \frac{\delta W({a})}{\delta a} + \tau_{1} \sigma \xi,
\end{equation} which corresponds to a damped nonlinear wave equation with additive space-white noise.

%
%
%
%

\noindent{\bf Second order model.} When $k=2$, by introducing auxiliary variables ${z}_{1}(=z)$ and ${z}_{2}$, we get
\begin{equation}\begin{aligned}
&\dot{{z}}_{1} = A_{1}{b}+B_{1}{z}_{1}+{z}_{2}, 
&\dot{{z}}_{2} = A_{2}{b}+B_{2}{z}_{1}.\\
\end{aligned}
\label{eq: model2}
\end{equation}

We can write the second order model with noise in a more compact form, 
$ \dot{{z}}=  A b + B{z} + \sigma \xi$.
with   $A$ and $B$ being block matrices,
\[
A=\left[ \begin{array}{c}
A_{1} \\ A_{2}
\end{array}
\right]
 \text{ and } 
B= \left[
\begin{array}{cc}
B_{1} & I \\
B_{2} & 0\\
\end{array}
\right].
\]
Then the equation  has an equilibrium density, $\rho(a,z)\propto \exp -\left(W({a}) + \frac12 {z}^\intercal Q^{-1} {z}\right) $ by choosing $Q$ according to $\sigma \sigma^{\intercal} = -(BQ + QB^{\intercal})$. 


From numerical observations, we have $M_{4}\approx -\gamma_{2}\nabla_{h}^{2}, N_{2}\approx -\alpha \nabla_{h}^{2}$ and $\gamma_{2}\gg \gamma_{1}\alpha$. So we have,
\begin{equation}\left\{\begin{aligned}
&\dot{{a}} = {z}_{1}, \\
&\dot{{z}}_{1} =  \nabla_h^2   \gamma_1 \frac{\delta W({a})}{\delta a}  -\frac{\gamma_2\kappa}{\gamma_1^{2}}{z}_{1}+{z}_{2} , \\
&\dot{{z}}_{2} = \nabla_h^2   \frac{\gamma_2\kappa}{\gamma_{1}} \frac{\delta W({a})}{\delta a} -\frac{\gamma_2}{\gamma_{1}} {z}_{1}+ \sigma \xi(t),\\
\end{aligned}
\label{eq: secondmodel}\right.
\end{equation}
where $\sigma\sigma^{\intercal} = -2 \frac{\gamma_{2}^{2}\kappa}{\gamma_{1}^{2}}\nabla_{h}^{2}.$ 

\begin{figure}[H]
\begin{center}
\includegraphics[width=0.38\textwidth]{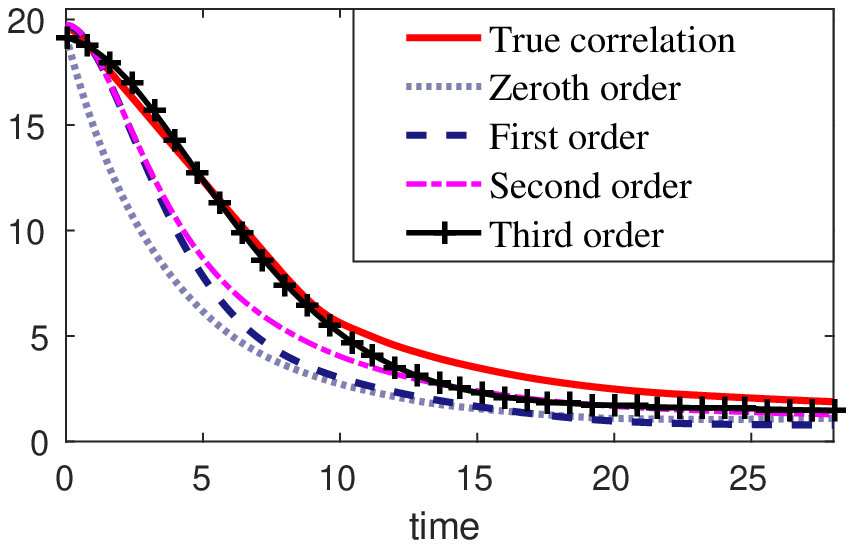}
\includegraphics[width=0.38\textwidth]{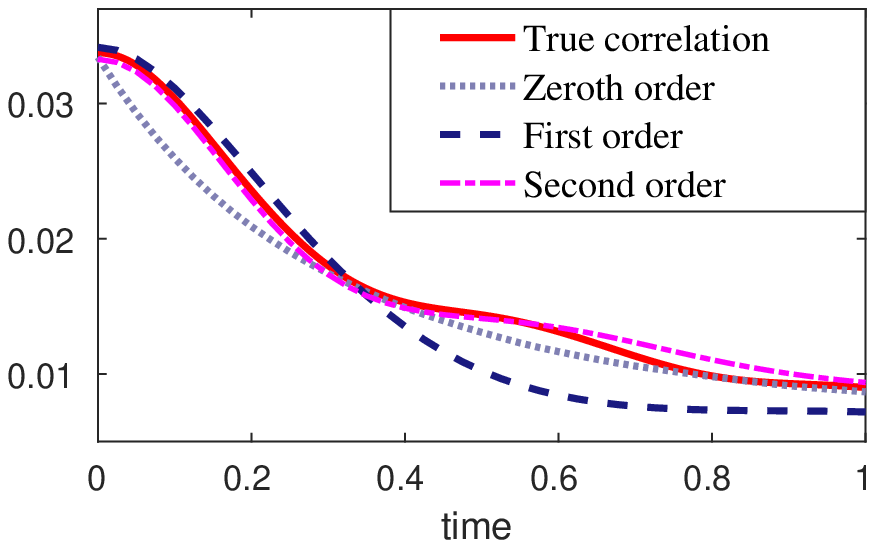}
\caption{Comparisons of true two-point statistics of $a_{1}$ with those from the reduced models. 
}
\label{fig: 1d-pdfcorr}
\end{center}
\end{figure}

Define the second relaxation parameter $\tau_{2}=\gamma_{1}/\gamma_{2}$. The corresponding constitutive relation for the heat flux $q$ can be written as,
\[
\begin{aligned}
\tau_{2} \ddot{q} + \tau_{1}\dot{q} + q & = -\nabla_h  \kappa \frac{\delta W({a})}{\delta a}-\frac{d}{dt} \nabla  \frac{\tau_{2}\kappa}{\tau_{1}} \frac{\delta W({a})}{\delta a}\\ &+ \sqrt{2\kappa} \xi(t). 
\end{aligned}
\]
We compare the numerical results from the stochastic models \eqref{eq: zero_GLE}, \eqref{eq: firstmodel}, 
\eqref{eq: secondmodel}, and a third order model, to the statistics obtained from full MD simulations. As shown in Figure \ref{fig: 1d-pdfcorr}.
the two-point statistics of the energy is consistent for both the 1D chain model (top) and the nanotube system (bottom). We  observed improved accuracy as we increase the order of the Markovian embedding approximations. 
In summary, we derived rigorously generalized heat conduction models  from the underlying MD model. The stochastic constitutive equation can be nonlinear and the parameters were linked directly to the statistics of local energy, making it possible to determine system-specific model parameters. The models were validated by examining the two-point statistics. 

%

\end{document}